\documentclass[conference]{IEEEtran}
\IEEEoverridecommandlockouts

\usepackage{graphicx}
\usepackage{caption}  
\usepackage{subcaption}  
\usepackage{cite}
\usepackage{amsmath,amssymb,amsfonts}
\usepackage{algorithmic}
\usepackage{textcomp}
\usepackage{xcolor}
\usepackage{color}
\usepackage{psfrag}
\usepackage{epsfig}
\usepackage{enumerate}
\usepackage[top=0.71in, bottom=1in, left=0.625in, right=0.625in]{geometry}
\begin{document}

\title{Green Learning for STAR-RIS mmWave Systems with Implicit CSI \vspace{-0.1in}
\thanks{This work was supported in part by the Academia Sinica (AS) under Grant 235g Postdoctoral Scholar Program,  the Sixth Generation Communication and Sensing Research Center funded by the Higher Education SPROUT Project, the Ministry of Education of Taiwan, and the National Science and Technology Council (NSTC) of Taiwan under Grant 113-2221-E-110-059-MY3, 113-2218-E-110-008, 113-2218-E-110-009, and 113-2926-I-001-502-G. Prof. Walid Saad was supported by the U.S. National Science Foundation (NSF) under Grant CNS-2225511.}
}

\author{\IEEEauthorblockN{Yu-Hsiang Huang$^{1}$, Po-Heng Chou$^{2}$, Wan-Jen Huang$^{1}$, Walid Saad$^{3}$, and C.-C. Jay Kuo$^{4}$}
\IEEEauthorblockA{$^{1}$Institute of Communication Engineering (ICE), National Sun Yat-sen University (NSYSU), Kaohsiung 80424, Taiwan\\
$^{2}$Research Center for Information Technology Innovation (CITI), Academia Sinica (AS), Taipei 11529, Taiwan\\
$^{3}$Bradley Department of Electrical and Computer Engineering (ECE), Virginia Tech (VT), Alexandria, VA 22305, USA\\
$^{4}$Ming Hsieh Department of Electrical Engineering, University of Southern California (USC), CA 90089, USA\\
\small E-mails: m123070003@nsysu.edu.tw, d00942015@ntu.edu.tw, wjhuang@faculty.nsysu.edu.tw, walids@vt.edu, cckuo@sipi.usc.edu}
\vspace{-0.25in}
}
\maketitle

\begin{abstract}
In this paper, a green learning (GL)-based precoding framework is proposed for simultaneously transmitting and reflecting reconfigurable intelligent surface (STAR-RIS)-aided millimeter-wave (mmWave) MIMO broadcasting systems. Motivated by the growing emphasis on environmental sustainability in future 6G networks, this work adopts a broadcasting transmission architecture for scenarios where multiple users share identical information, improving spectral efficiency and reducing redundant transmissions and power consumption. Different from conventional optimization methods, such as block coordinate descent (BCD) that require perfect channel state information (CSI) and iterative computation, the proposed GL framework operates directly on received uplink pilot signals without explicit CSI estimation. Unlike deep learning (DL) approaches that require CSI-based labels for training, the proposed GL approach also avoids deep neural networks and backpropagation, leading to a more lightweight design. Although the proposed GL framework is trained with supervision generated by BCD under full CSI, inference is performed in a fully CSI-free manner. The proposed GL integrates subspace approximation with adjusted bias (Saab), relevant feature test (RFT)-based supervised feature selection, and eXtreme gradient boosting (XGBoost)-based decision learning to jointly predict the STAR-RIS coefficients and transmit precoder. Simulation results show that the proposed GL approach achieves competitive spectral efficiency compared to BCD and DL-based models, while reducing floating-point operations (FLOPs) by over four orders of magnitude. These advantages make the proposed GL approach highly suitable for real-time deployment in energy- and hardware-constrained broadcasting scenarios.
\end{abstract}

\section{Introduction}


Reconfigurable intelligent surfaces (RISs) are a key enabler for addressing the severe path loss, limited diffraction, and blockage sensitivity of millimeter-wave (mmWave) communication in next-generation wireless systems~\cite{Changsheng2025}. By enabling programmable control of the propagation environment through nearly-passive meta-surfaces, RISs support passive beamforming, virtual line-of-sight links, and energy-efficient coverage extension in MIMO systems~\cite{Chou2024IWCL, Chou2024Globecom, Huang2024PIMRC}. However, conventional RISs are limited to one-sided reflection, which restricts full-space coverage, especially in 3D network deployments with users on both sides. To overcome this, the concept of simultaneously transmitting and reflecting RIS (STAR-RIS) has been proposed~\cite{Xidong2022}, allowing each element to independently modulate both reflected and transmitted signals. This enables two-sided coverage, enhances spatial design flexibility, and promotes energy-efficient, full-space wireless networking.


To implement simultaneous transmission and reflection in practice, three operating protocols for STAR-RIS can be considered: Energy splitting (ES), mode switching (MS), and time switching (TS)~\cite{Changsheng2025, Xidong2022}. In the ES protocol, each element simultaneously handles both transmission and reflection by allocating a controllable power ratio between the two directions. While this mode provides the highest spatial flexibility and beamforming granularity, it also introduces a large number of coupled optimization variables, increasing system complexity. In contrast, the MS protocol partitions the RIS elements into two disjoint sets, where each element is dedicated to either transmission or reflection at a given time, thereby reducing control complexity at the expense of some beamforming flexibility. The TS protocol further simplifies implementation by alternating the entire surface between transmission and reflection phases over time, which decouples the optimization of the two functions but requires strict time synchronization. 



\textcolor{black}{
To improve content delivery efficiency and support massive IoT and URLLC in 6G communication systems~\cite{Changsheng2025}, broadcasting transmission has gained increasing attention for multi-user scenarios where identical information must be delivered to multiple receivers~\cite{Fallgren2019TBC}. In autonomous driving networks, for example, shared information such as road conditions needs to be distributed to a fleet of vehicles that may dynamically enter or leave the coverage area. Unlike unicast, which repeatedly transmits the same data stream to each user, broadcasting delivers a common signal to multiple users simultaneously, reducing redundant transmissions, lowering power consumption, and improving spectral efficiency~\cite{Du2021TWC}. Motivated by these benefits, this work considers a STAR-RIS-assisted MIMO broadcasting system operating under the ES protocol, providing a practical solution for low-power, full-space communication in 6G networks. Recent studies have investigated precoding strategies for STAR-RIS-assisted MIMO systems using optimization-based and deep learning (DL)-based methods~\cite{Niu2022TVT, Zhong2022JSAC, Perdana2025TWC, Sun2025CL}. Classical optimization techniques typically rely on full CSI and adopt alternating optimization to jointly design the precoder and transmission/reflection coefficients (TRCs), as exemplified by the block coordinate descent (BCD) approach in~\cite{Niu2022TVT}.}

\textcolor{black}{Building on the above optimization-based approaches, to overcome the limitations of conventional methods that rely on full CSI, DL-based solutions have been proposed.} For instance, hybrid deep reinforcement learning has been used to handle discrete TRC selection under coupled constraints~\cite{Zhong2022JSAC}, and convolutional neural networks (CNNs) have been applied to approximate STAR-RIS configurations in short-packet simultaneous wireless information and power transfer non-orthogonal multiple access (SWIPT-NOMA) systems~\cite{Perdana2025TWC}. However, these methods still rely on large-scale deep neural networks (DNNs), demand full or high-fidelity CSI labels for training, and require substantial inference resources at runtime.
Beyond full CSI designs, some recent efforts have attempted to relax CSI dependency. For example, the authors in~\cite{Sun2025CL} proposed a statistical CSI-based design that integrates STAR-RIS with movable antennas, jointly optimizing beamforming, TRCs, and antenna positions using fractional programming. While this approach partially relaxes the CSI assumption, it still depends on iterative solvers and does not leverage pilot-based or learning-based inference. Nevertheless, no existing method has yet leveraged uplink pilots for direct, CSI-free precoding inference.

To the best of our knowledge, no prior work has directly leveraged uplink pilot signals to infer STAR-RIS precoding decisions without relying on explicit CSI. Meanwhile, the potential of lightweight and interpretable learning frameworks, green learning (GL) proposed in~\cite{Kuo2023IntroGL}, remains largely unexplored in this context. Unlike conventional neural networks that rely on backpropagation, the GL framework adopts a feedforward, layer-wise design, where each stage is constructed using lightweight processing modules. This architecture significantly reduces model complexity and training overhead, offering a practical balance between inference performance and resource efficiency~\cite{Kuo2023IntroGL, Kuo2019Saab, Yang2022RFT}. These properties make GL particularly suitable for mmWave systems operating under energy and hardware constraints. GL has also been applied to conventional RIS-aided systems~\cite{Liao2024PIMRC}, demonstrating its effectiveness in CSI-free precoding based on pilot signals. However, these works were limited to one-sided reflection scenarios and did not address the unique challenges of STAR-RIS.

The main contribution of this paper is, thus, a GL-based precoding framework for STAR-RIS-assisted \textcolor{black}{broadcasting} MIMO systems under the ES protocol. The proposed GL approach operates directly on received uplink pilot signals and comprises three stages: (i) subspace approximation with adjusted bias (Saab)~\cite{Kuo2019Saab} for unsupervised feature extraction, (ii) relevant feature test (RFT)~\cite{Yang2022RFT} for supervised feature selection, and (iii) eXtreme gradient boosting (XGBoost)~\cite{Chen2016XGBoost} for interpretable decision inference. While the GL model requires no CSI during inference, its supervised feature selection uses optimal precoding labels generated by the BCD algorithm under perfect CSI. This setup allows GL to mimic near-optimal behavior at runtime without relying on CSI estimation. This architecture eliminates the need for DNNs, yielding ultra-low floating-point operations (FLOPs) and enabling real-time deployment on constrained hardware.

In summary, our key contributions include:
\begin{itemize}
    \item We propose a GL-based precoding framework for STAR-RIS-assisted mmWave \textcolor{black}{broadcasting} MIMO systems, which operates directly on received uplink pilot signals without requiring CSI during inference.
    \item Unlike conventional optimization methods such as BCD that require full CSI and iterative updates, the proposed GL learns a data-driven mapping from pilots to precoding decisions through a lightweight, feedforward architecture.
    \item While supervised feature selection of the proposed GL approaches by BCD under perfect CSI, the proposed GL framework achieves fully CSI-free inference.
    \item Simulation results demonstrate that GL achieves competitive spectral efficiency compared to BCD and DL-based models, while reducing FLOPs by over four orders of magnitude, making it suitable for energy- and hardware-constrained deployment.
\end{itemize}

\section{System Model and Channel Model}

\begin{figure}[t]
    \centering
    \begin{subfigure}[b]{0.4\textwidth}
        \includegraphics[width=\textwidth]{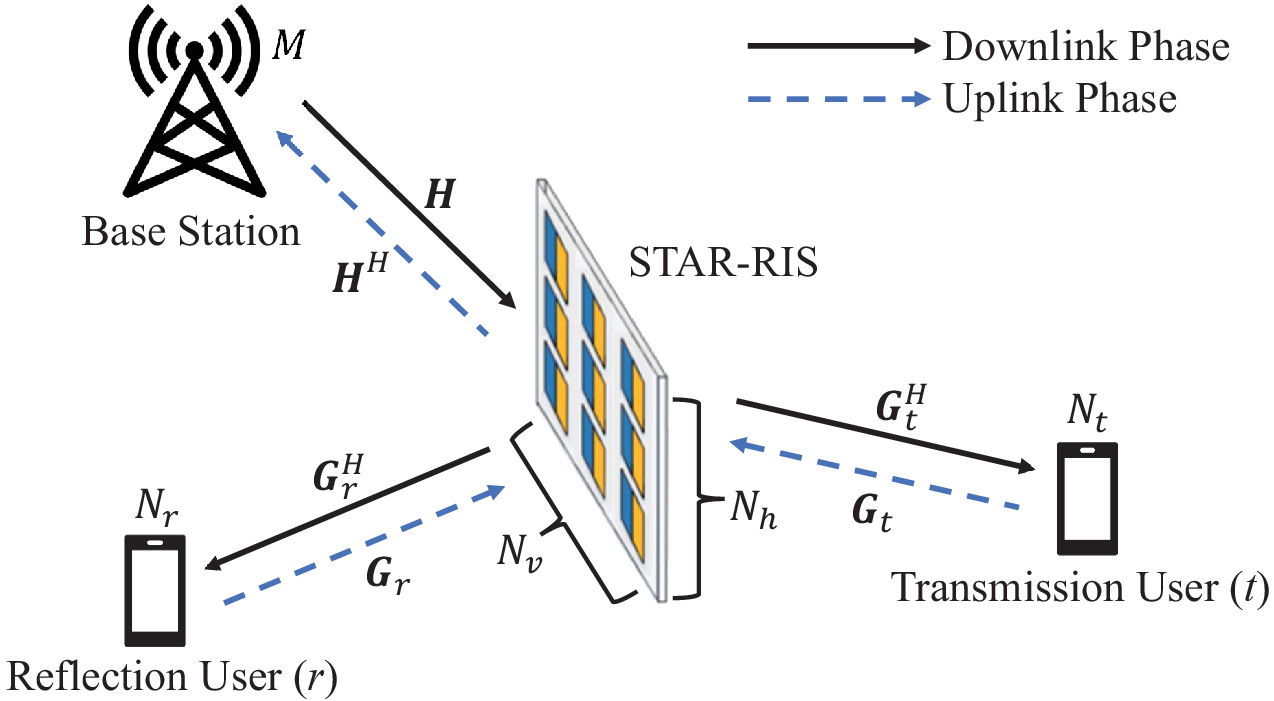}
    \end{subfigure}
    \captionsetup{font=small} 
    \caption{The STAR-RIS-aided MIMO system.}\label{fig:system_model} 
    \vspace{-0.2in}
\end{figure}

We consider a STAR-RIS-aided mmWave MIMO system consisting of a base station (BS) equipped with $M$ antennas, and two users: a reflection user ($r$) and a transmission user ($t$), each equipped with $N_r$ and $N_t$ antennas, respectively. A uniform planar array (UPA)-based STAR-RIS with $N_h \times N_v = N$ passive elements is deployed between the BS and the users, as illustrated in Fig.~\ref{fig:system_model}.

\textcolor{black}{
During the downlink phase, denote the $n$-th unit-energy broadcasting symbol transmitted by the BS as $x[n]$. The signal received by user $l \in \{t, r\}$ is given by }
\begin{equation}
    y_l[n] = \boldsymbol{G}_l^H \boldsymbol{\Phi}_l \boldsymbol{H} \textcolor{black}{\boldsymbol{w}} \boldsymbol{x}[n] + \textcolor{black}{\boldsymbol v_{l}[n]},
\end{equation}
where \textcolor{black}{$\boldsymbol{w}\! \in \!\mathbb{C}^{M \times 1}$} is the precoding vector at the BS, $\boldsymbol{H} \!\in \!\mathbb{C}^{N\! \times \!M}$ is the channel from the BS to the STAR-RIS, $\boldsymbol{G}_l^H \!\in \!\mathbb{C}^{N_l \times N}$ is the channel from the STAR-RIS to user $l$, $\boldsymbol{\Phi}_l \!\in\! \mathbb{C}^{N \times N}$ is the STAR-RIS configuration matrix, and \textcolor{black}{$\boldsymbol v_{l}[n]$} $\sim \mathcal{CN}(0, \sigma^2 \boldsymbol{I})$ represents additive white Gaussian noise (AWGN). In this work, it is assumed that the channel reciprocity holds~\cite{Jiang2021}. The STAR-RIS configuration $\boldsymbol{\Phi}_l$ is defined as
\begin{equation}
    \boldsymbol{\Phi}_l = \operatorname{diag}(\alpha_{l,1} e^{j\theta_{l,1}}, \dots, \alpha_{l,N} e^{j\theta_{l,N}}),
\end{equation}
where $\alpha_{l} \in (0, 1]$ is the amplitude response that satisfy $\alpha_{r}^2 + \alpha_{t}^2 = 1$, and $\theta_{l,n} \in [0, 2\pi)$ is the phase shift of the $n$-th element that $\left| e^{j\theta_{l,N}} \right| = 1$. \textcolor{black}{The BS precoding vector $\boldsymbol{w}$} satisfies the power constraint \textcolor{black}{$\|\boldsymbol{w}\|_F^2 \leq P_t$}.

During the uplink phase, both users transmit pilot signals $\boldsymbol{p}_l \!\in \!\mathbb{C}^{N_l\! \times\! 1}$, and the received pilot signal at the BS is written by
\begin{equation}
    \boldsymbol{y}[n] = \sum_{l \in \{t, r\}} \boldsymbol{H}^H \boldsymbol{\Phi}_l \boldsymbol{G}_l \boldsymbol{p}_l[n] + \textcolor{black}{\boldsymbol{v}[n]}.
\end{equation}
Due to reciprocity, the downlink CSI is implicitly embedded in the uplink pilot response.

We assume Rician fading for all links, where the BS-to-STAR-RIS channel $\boldsymbol{H}$ and STAR-RIS-to-user ($l$) channels $\boldsymbol{G}_l^H$ ($l \in \{r,t\}$) are modeled as
\begin{align}
\boldsymbol{H} &= \sqrt{\beta_q} \left( \sqrt{\frac{K}{K+1}} \bar{\boldsymbol{H}}_q + \sqrt{\frac{1}{K+1}} \tilde{\boldsymbol{H}}_q \right), \\
\boldsymbol{G}_l^H &= \sqrt{\beta_l} \left( \sqrt{\frac{K}{K+1}} \bar{\boldsymbol{G}}_l^H + \sqrt{\frac{1}{K+1}} \tilde{\boldsymbol{G}}_l^H \right),
\end{align}
where $\beta_q$, $\beta_r$, and $\beta_t$ represent the path loss coefficients for the respective links, and $K$ is the Rician factor. All antenna elements are placed with half-wavelength spacing.

The STAR-RIS steering vector is defined as
\begin{equation}
\boldsymbol{a}_{\mathrm{RIS}}(\phi, \varphi) = \boldsymbol{a}_{N_v}(\sin\varphi) \otimes \boldsymbol{a}_{N_h}(\cos\varphi \sin\phi),
\end{equation}
where $\boldsymbol{a}_N(\psi) = [1, e^{j\pi \psi}, \dots, e^{j(N-1)\pi \psi}]$. 

The line-of-sight (LOS) components of the BS-to-STAR-RIS and STAR-RIS-to-user ($l$) links are given by
\begin{align}
\bar{\boldsymbol{H}}_q &= \boldsymbol{a}_{\mathrm{RIS}}^H(\phi_{A,0}, \varphi_{A,0}) \otimes \boldsymbol{a}_M(\phi_{D,0}^{(q)}), \\
\bar{\boldsymbol{G}}_l^H &= \boldsymbol{a}_{N_l}^H(\phi_{A,0}^{(l)}) \otimes \boldsymbol{a}_{\mathrm{RIS}}(\phi_{D,0}, \varphi_{D,0}),
\end{align}
where the angles $\phi_{A,0}, \varphi_{A,0}$ and $\phi_{D,0}, \varphi_{D,0}$ represent the elevation and azimuth angles of arrival and departure at the STAR-RIS, respectively, while $\phi_{D,0}^{(q)}$ and $\phi_{A,0}^{(l)}$ represent the angle of departure (AOD) at the BS and the angle of arrival (AOA) at the user side, respectively.

The non-line-of-sight (NLOS) components of the BS-to-STAR-RIS and STAR-RIS-to-user ($l$) links are expressed as
\begin{align}
\tilde{\boldsymbol{H}}_q &= \sum_{\ell=1}^{L_q} z_{q,\ell} \boldsymbol{a}_{\mathrm{RIS},\ell}^H \otimes \boldsymbol{a}_M(\phi_{D,\ell}^{(q)}), \\
\tilde{\boldsymbol{G}}_l^H &= \sum_{\ell=1}^{L_l} z_{l,\ell} \boldsymbol{a}_{N_l}^H(\phi_{A,\ell}^{(l)}) \otimes \boldsymbol{a}_{\mathrm{RIS},\ell},
\end{align}
where $z_{q,\ell}, z_{l,\ell} \sim \mathcal{CN}(0, 1/L_l)$ are the complex path gains, and $L_q$, $L_l$ represent the number of multipaths for the BS-to-RIS and RIS-to-user links, respectively. Similarly, $\phi_{A,\ell}^{(l)}$ and $\phi_{D,\ell}^{(q)}$ represent the AOA and AOD for the $\ell$-th NLOS path at the user $l$ and BS, respectively.

\section{Problem Formulation}

Based on the system model described in the previous section, the effective channel between the BS and user $l \in \{t,r\}$ is given by
\begin{equation}
    \boldsymbol{H}_{\mathrm{eff},l} = \boldsymbol{G}_l^H \boldsymbol{\Phi}_l \boldsymbol{H},
\end{equation}
where $\boldsymbol{H}_{\mathrm{eff},l} \in \mathbb{C}^{N_l \times M}$ represents the cascaded BS-STAR-RIS-user channel. 

Our objective is to jointly optimize the BS \textcolor{black}{precoding vector $\boldsymbol{w}$} and the STAR-RIS configuration matrix $\boldsymbol{\Phi}_l$ to maximize the achievable rate of user $l$, \textcolor{black}{since our work adopts a broadcasting transmission architecture, interference from other users is not considered.}
\begin{align}
\max_{\textcolor{black}{\boldsymbol{w}}, \boldsymbol{\Phi}_l} \quad & R_l = \log_2 \left| \boldsymbol{I}_{N_s} + \frac{1}{\sigma^2} \boldsymbol{H}_{\mathrm{eff},l} \textcolor{black}{\boldsymbol{w}} \textcolor{black}{\boldsymbol{w}}^H \boldsymbol{H}_{\mathrm{eff},l}^H \right|, \tag{12a} \label{eq:rate_obj} \\
\text{s.t.} \quad & \boldsymbol{\Phi}_l = \operatorname{diag}(\alpha_{l,n} e^{j\theta_{l,n}}), \tag{12b} \label{constraint:phi} \\
& \alpha_{l,n} \in (0,1], \quad \theta_{l,n} \in [0,2\pi), \tag{12c} \label{constraint:alpha_theta} \\
& \|\textcolor{black}{\boldsymbol{w}}\|_F^2 \leq P_t, \tag{12d} \label{constraint:power}
\end{align}
where $|\cdot|$ represents the matrix determinant. Constraints~\eqref{constraint:phi} and~\eqref{constraint:alpha_theta} ensure that each STAR-RIS element adheres to the amplitude-phase response under the ES protocol, while constraint~\eqref{constraint:power} limits the total transmission power at the BS.

Existing research has explored the optimization of STAR-RIS using classical algorithms~\cite{Niu2022TVT}. In particular, the weighted minimum mean square error (WMMSE) method introduces auxiliary variables to decompose the original problem into multiple subproblems, which are then solved via BCD~\cite{Razaviyayn2013BCD} to maximize the downlink achievable rate. However, these approaches rely on perfect CSI and assume fully passive reflective elements, resulting in high overhead for channel estimation between the BS, the STAR-RIS, and the users.

To overcome these limitations, next, we propose a data-driven GL framework~\cite{Kuo2023IntroGL} that bypasses explicit CSI estimation by directly operating on the received pilot signals. The proposed GL approach aims to jointly predict the optimal $\boldsymbol{w}$ and $\boldsymbol{\Phi}_l$ configurations in a low-complexity, CSI-free manner.

\section{Proposed Green Learning (GL) Framework}

To solve the precoding problem in \eqref{eq:rate_obj} using only uplink pilot signals, we propose a lightweight and interpretable GL framework. This method consists of four main stages: (i) pilot signal construction, (ii) unsupervised representation learning via Saab, (iii) supervised feature selection via RFT, and (iv) decision learning via XGBoost. These modules collectively transform raw pilot measurements into near-optimal precoding decisions with minimal complexity.

\subsection{Stage 1: Pilot Signal Construction}
To enable efficient exploration of channel status, we develop a hierarchical pilot structure that incorporates the pilot signals transmitted from users and the phase-shifting and amplitude-scale patterns at the RIS, as illustrated in Table~\ref{table:pilot}. Let $\boldsymbol{P} = [\boldsymbol{p}_1, \dots, \boldsymbol{p}_{N_p}] \in \mathbb{C}^{(N_r + N_t)) \times N_p}$ be a set of full-rank pilot vectors. In this work, the matrix $\boldsymbol{P}$ is constructed from orthogonal space-time block codes such that the rows in $\boldsymbol{P}$ are mutually orthogonal.

While user $r$ and $t$ repeatedly transmit pilot vectors $\boldsymbol{p}_i\in \mathbb{C}^{(N_r + N_t)) \times 1}$, each pilot vector corresponds to $N$ distinct STAR-RIS phase shift response matrices $\boldsymbol{\Theta}_l \in \mathbb{C}^{N\times N} = \operatorname{diag}\left(e^{j\theta_{l,1}},\, e^{j\theta_{l,2}},\, \dots,\, e^{j\theta_{l,N}}\right) = [\boldsymbol{\Theta}_{l,1}, \boldsymbol{\Theta}_{l,2}, \dots, \boldsymbol{\Theta}_{l,N}]$.
At the STAR-RIS, the phase shift vectors are chosen from a discrete Fourier transform (DFT)-based codebook $\mathcal{C}$~\cite{Yang2010}, which is defined as
\begin{equation}
    \mathcal{C} = \left\{ \vartheta \left( N_h, \tfrac{2\pi n_h}{N_h} \right) \otimes \vartheta \left( N_v, \tfrac{2\pi n_v}{N_v} \right) \right\},
\end{equation}
where $n_h = 0, 1, \dots, N_h - 1$, $n_v = 0, 1, \dots, N_v - 1$, and $\vartheta(N,\theta) = [1, e^{-j\theta}, \dots, e^{-j(N-1)\theta}]$.
Furthermore, each $\boldsymbol{\Theta}_{l,j}$ is associated with a different amplitude-scaling matrix.
Each amplitude-scaling matrix $\boldsymbol{\alpha}_{l,k} \in \mathbb{R}^{N \times N}$ is defined as $\boldsymbol{\alpha}_{l,k} = \operatorname{diag}(\alpha_{l,k,1}, \dots, \alpha_{l,k,N})$, where the factor $\alpha_{l,k,n}$ is chosen from a preset finite set.
These control patterns are systematically paired with each pilot vector $\boldsymbol{p}_i$ to generate a total of $N_p \times N \times K$ received signal samples $\{\boldsymbol{y}_1, \boldsymbol{y}_2, \dots, \boldsymbol{y}_{N_p N K} \}$.

\begin{table}[t]
\centering
\caption{Hierarchical mapping from pilot to received signal.}
\label{table:pilot}
\renewcommand{\arraystretch}{1.05}
\setlength{\tabcolsep}{5pt}
\begin{tabular}{|c|c|c|c|}
\hline
\textbf{Pilot} & \textbf{Phase} & \textbf{Amplitude} & \textbf{Received signal} \\
\hline
$\boldsymbol{p}_{1}$ & $\boldsymbol{\Theta}_{l,1}$ & $\boldsymbol{\alpha}_{l,1}$ & $\boldsymbol{y}_{1}$ \\
                     &                               & $\vdots$                  & $\vdots$ \\
                     &                               & $\boldsymbol{\alpha}_{l,K}$ & $\boldsymbol{y}_{K}$ \\
\cline{2-4}
                     & $\boldsymbol{\Theta}_{l,2}$   & $\boldsymbol{\alpha}_{l,1}$ & $\boldsymbol{y}_{K+1}$ \\
                     &                               & $\vdots$                  & $\vdots$ \\
                     &                               & $\boldsymbol{\alpha}_{l,K}$ & $\boldsymbol{y}_{2K}$ \\
\cline{2-4}
                     & $\vdots$                      & $\vdots$                  & $\vdots$ \\
\cline{2-4}
                     & $\boldsymbol{\Theta}_{l,N}$   & $\boldsymbol{\alpha}_{l,1}$ & $\boldsymbol{y}_{NK-K+1}$ \\
                     &                               & $\vdots$                  & $\vdots$ \\
                     &                               & $\boldsymbol{\alpha}_{l,K}$ & $\boldsymbol{y}_{NK}$ \\
\hline
$\boldsymbol{p}_{2}$ & $\boldsymbol{\Theta}_{l,1}$   & $\boldsymbol{\alpha}_{l,1}$ & $\boldsymbol{y}_{NK+1}$ \\
                     & $\vdots$                      & $\vdots$                  & $\vdots$ \\
                     & $\boldsymbol{\Theta}_{l,N}$   & $\boldsymbol{\alpha}_{l,K}$ & $\boldsymbol{y}_{2NK}$ \\
\hline
$\vdots$             & $\vdots$                      & $\vdots$                  & $\vdots$ \\
\hline
$\boldsymbol{p}_{N_p}$ & $\boldsymbol{\Theta}_{l,1}$ & $\boldsymbol{\alpha}_{l,1}$ & $\boldsymbol{y}_{(N_p-1)NK+1}$ \\
                       & $\vdots$                    & $\vdots$                  & $\vdots$ \\
                       & $\boldsymbol{\Theta}_{l,N}$ & $\boldsymbol{\alpha}_{l,K}$ & $\boldsymbol{y}_{N_p NK}$ \\
\hline
\end{tabular}
\vspace{-0.1in}
\end{table}

\subsection{Stage 2: Unsupervised Representation Learning}
To enable green learning, we first organize the received pilot signals of dimensions $N_p \times N \times K$ into a four-dimensional tensor $\boldsymbol{R} \in \mathbb{C}^{M \times N_p \times N \times K}$. The compact and informative representations are then extracted from $\boldsymbol{R}$ using a four-stage Saab transformation~\cite{Kuo2019Saab}. In the first stage, the tensor is partitioned into $N_p \times N \times K$ patches of size $M$, and each patch is projected onto a set of antenna-domain anchor vectors. This projection yields a $N_p \times N \times K$ cube for each anchor vector. In the subsequent stages, these $M$ cubes are further projected along the amplitude-scaling, phase-shifting, and pilot-signal domains, corresponding to the second, third, and fourth stages of the transformation, respectively. Each stage utilizes a subspace composed of a DC anchor vector and multiple AC anchor vectors, where the AC anchors are derived from principal component analysis (PCA) after mean removal. At each stage, components with negligible average energy are discarded, resulting in a compact and decorrelated set of features.

\subsection{Stage 3: Supervised Feature Selection}
While Saab reduces the dimensionality of the received pilot signals, not all extracted features are informative. To further refine the input space, we adopt the relevant feature test (RFT)~\cite{Yang2022RFT}, a supervised feature selection method tailored for regression tasks.

Let $z_f$ represent the $f$-th Saab coefficient, and let $t$ be a threshold scanned uniformly over the feature range $[t_{\min}^{(f)}, t_{\max}^{(f)}]$. For each candidate threshold, the labeled dataset is split into left and right segments based on the value of $z_f$. The predicted label in each segment is taken as the sample mean, and the weighted mean square error (WMSE) between the predicted and true labels is computed.
The relevance score for $z_f$ is defined as the minimal WMSE across all thresholds
\begin{equation}
    \mathcal{L}_{\text{RFT}}^{(f)} = \min_{t \in (t_{\min}^{(f)}, t_{\max}^{(f)})} \text{WMSE}(z_f, t).
\end{equation}
Features with smaller $\mathcal{L}_{\text{RFT}}^{(f)}$ are considered more predictive and are selected as the final input for the regression stage.
The regression targets required by RFT are generated offline using BCD under perfect CSI, enabling a supervised feature selection process without relying on CSI during inference.

\subsection{Stage 4: Supervised Decision Learning}

After RFT-based selection, the most relevant features are used to train a regression model based on XGBoost~\cite{Chen2016XGBoost}. XGBoost constructs an ensemble of decision trees in a forward stage-wise fashion, allowing each new tree to correct errors from previous ones. This method enables efficient learning of complex nonlinear mappings between selected features and the optimal precoding decisions.

The regression targets include the phase shifts of the STAR-RIS and the linear precoder at the BS. The phase shifts are represented using a trigonometric embedding as
\begin{equation}
    \boldsymbol{u}_\Phi = [\cos \theta_1, \dots, \cos \theta_N, \sin \theta_1, \dots, \sin \theta_N],
\end{equation}
where $\theta_n$ represents the optimal phase shift for the $n$-th STAR-RIS element. The BS precoder is vectorized as \textcolor{black}{$\boldsymbol{u}_w = \text{vec}(\boldsymbol{w})$, and the complete regression target is $[\boldsymbol{u}_\Phi, \boldsymbol{u}_w]$}.

Compared to DL-based approaches, XGBoost does not require backpropagation or iterative weight updates. It also supports internal feature subsampling during training, which enhances generalization and reduces inference complexity. This makes it highly suitable for hardware-constrained deployments where model transparency and runtime efficiency are critical.


\section{Simulation Results}

In our simulations, we consider a STAR-RIS-aided mmWave MIMO system, where the BS is equipped with $M = 8$ antennas and the users with $N_r = N_t = {1, 4}$ antennas. The transmission power is varied over $P \in {10, 20, 30, 40, 50}$ dBm, and the noise variance is fixed at $\sigma^2 = -100$ dBm. The STAR-RIS is modeled as a UPA composed of $N = 16$ ($N_h = N_v = 4$) passive elements. The coordinates of the BS, STAR-RIS, reflection user $r$, and transmission user $t$ are set to $(0, 20, 0)$, $(0, 0, 0)$, $(5, 10, 0)$, and $(-5, -10, 0)$, respectively. For the channel model, the Rician factor is set to $K = 10$, with $L = 5$ NLOS paths per link. The path loss is modeled as $\beta = L (d/d_0)^{-\zeta}$, where $L = 10^{-1}$, $d_0 = 1$ meter, and $\zeta = 2$. AOAs and AODs are drawn uniformly from $[0, \pi]$. All results are averaged over 1000 random channel realizations.
 
To train the proposed GL framework, we adopt the BCD optimization algorithm~\cite{Niu2022TVT} under perfect CSI as a label generator. Specifically, BCD relies on accurate knowledge of the cascaded channels $\boldsymbol{H}$, $\boldsymbol{G}_r$, and $\boldsymbol{G}_t$ to jointly optimize the precoder $\boldsymbol{w}$ and the STAR-RIS coefficients $\boldsymbol{\Phi}_l$. While effective, such full CSI is rarely available in practice, and CSI estimation itself introduces significant computational and energy overhead.

To evaluate the effectiveness of our CSI-free GL-based approach, we compare against three baselines: (i) the conventional BCD algorithm under both ES and MS protocols~\cite{Niu2022TVT}; (ii) a fully-connected DNN with DFT-based codebook; and (iii) a Transformer~\cite{Transformer2017NeurIPS} with DFT-based codebook. Both the DNN and Transformer serve as representative DL-based approaches that operate without requiring explicit CSI at inference time, similar to the proposed GL scheme; (iv) Random selection.

For fair comparison, the DNN and Transformer models are trained using the same data and labels as GL. The DNN consists of four fully-connected layers with $1024$, $512$, $256$, and $128$ neurons, respectively, activated by Leaky ReLU, followed by an output layer using sigmoid or tanh activation. The Transformer adopts a typical encoder structure with eight attention heads, a positional embedding length of 16, and a feedforward network with $256$ neurons. Both models are trained using the Adam optimizer with a batch size of $256$, a learning rate of $10^{-4}$, and a training dataset of 100,000 samples. A validation set ($10\%$ of training data) is used for early stopping, and evaluation is performed on a separate test set of $10,000$ samples.
While DL-based approaches eliminate the need for CSI during inference, their large model size and heavy FLOPs demand may hinder real-time or edge deployment. In contrast, the proposed GL method achieves near-DNN performance with dramatically lower inference cost and no reliance on CSI at any stage.

Table~\ref{table:FLOPs} summarizes the computational cost of each scheme in terms of FLOPs per inference. 
For the $M = 8$, $N_r = 1$ configuration, the proposed \textcolor{black}{GL} method requires only \textcolor{black}{118,100} FLOPs, 
compared to \textcolor{black}{$13.97 \times 10^6$ and $8.12 \times 10^6$} FLOPs for the DNN and Transformer models, respectively. 
This corresponds to a complexity reduction of more than \textcolor{black}{two orders of magnitude}. 
A similar trend is observed in the $M = 8$, $N_r = 4$ setting, where \textcolor{black}{GL} requires only \textcolor{black}{491,100} FLOPs per inference. 
These results highlight the suitability of the proposed GL framework for energy- and hardware-constrained environments.

\begin{table}[t]
    \centering
    \caption{Comparison of FLOPs per inference.}
    \label{table:FLOPs}
    \setlength{\tabcolsep}{3pt} 
    \renewcommand{\arraystretch}{1} 
    \footnotesize 
    \begin{tabular}{|c|c|c|c|c|}
        \hline
        \textbf{Precoding schemes} & \textbf{BCD (ES)~\cite{Niu2022TVT}} & \textbf{\textcolor{black}{GL}} & \textbf{DNN} & \textbf{Transformer} \\ \hline
        $M=8$, $N_r=N_t=1$ & 7.16M & \textcolor{black}{0.1181M} & 13.97M & 8.12M \\ \hline
        $M=8$, $N_r=N_t=4$ & 7.86M & \textcolor{black}{0.4911M} & 26.58M & 28.41M \\ \hline
    \end{tabular}
\vspace{-0.2in}
\end{table}

\begin{figure}[t]
    \centering
    \begin{subfigure}[b]{0.31\textwidth}
        \includegraphics[width=\textwidth]{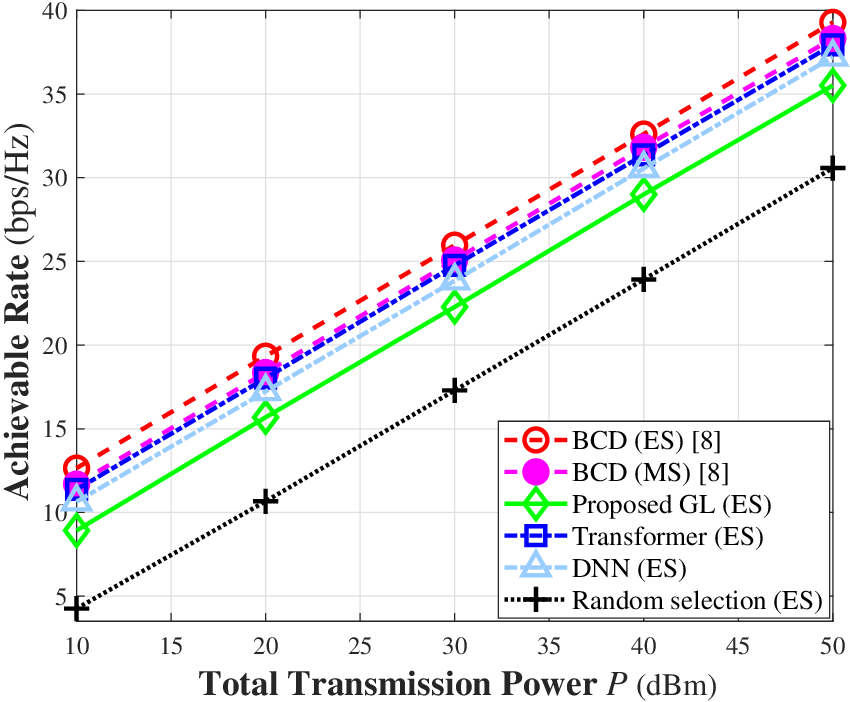}
        \caption{$M = 8$, $N_r = 1$}
        \label{fig:se_8x1}
    \end{subfigure}
    \par \vspace{0.1in}
    \begin{subfigure}[b]{0.31\textwidth}
        \includegraphics[width=\textwidth]{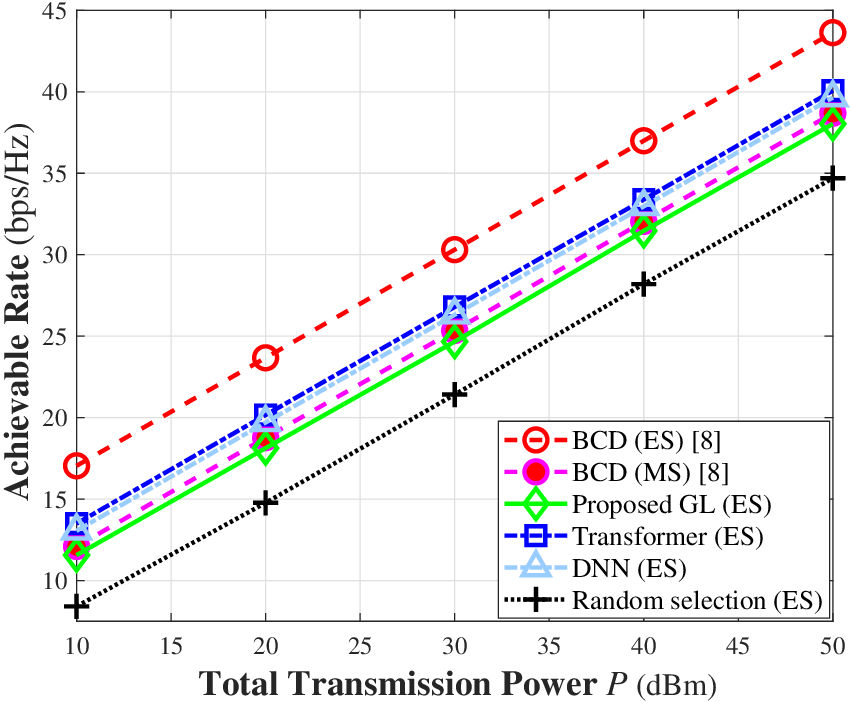}
        \caption{$M = 8$, $N_r = 4$}
        \label{fig:se_8x4}
    \end{subfigure}
    \captionsetup{font=small}
    \caption{Achievable rate versus transmission power under various precoding schemes.}
    \label{fig:main_power}
\vspace{-0.3in}
\end{figure}

Fig.~\ref{fig:se_8x1} presents the achievable rate versus transmission power under the $M = 8$, $N_r = N_t = 1$ configuration. The proposed GL closely follows the DNN and Transformer models, with an average rate gap of only $16.6\%$ and $5.7\%$ compared to the BCD (ES). At $P = 30$ dBm, GL achieves $85.8\%$ of the optimal BCD (ES) rate. Despite this modest gap, GL significantly reduces inference complexity, making it suitable for lightweight implementations. The random selection trails behind all other schemes with a gap exceeding $38\%$ on average.

Fig.~\ref{fig:se_8x4} compares the achievable rate under the $M = 8$, $N_r = N_t = 4$ configuration. As expected, the BCD (ES) yields the highest performance. The proposed GL remains within $20.4\%$ on average of the BCD (ES) and tracks within $6.5\%$ of the DNN and Transformer. Notably, GL achieves a rate of $24.7$ bps/Hz at $P=30$ dBm, demonstrating consistent behavior across spatially richer scenarios.

\begin{figure}[t]
    \centering
    \begin{subfigure}[b]{0.31\textwidth}
        \includegraphics[width=\textwidth]{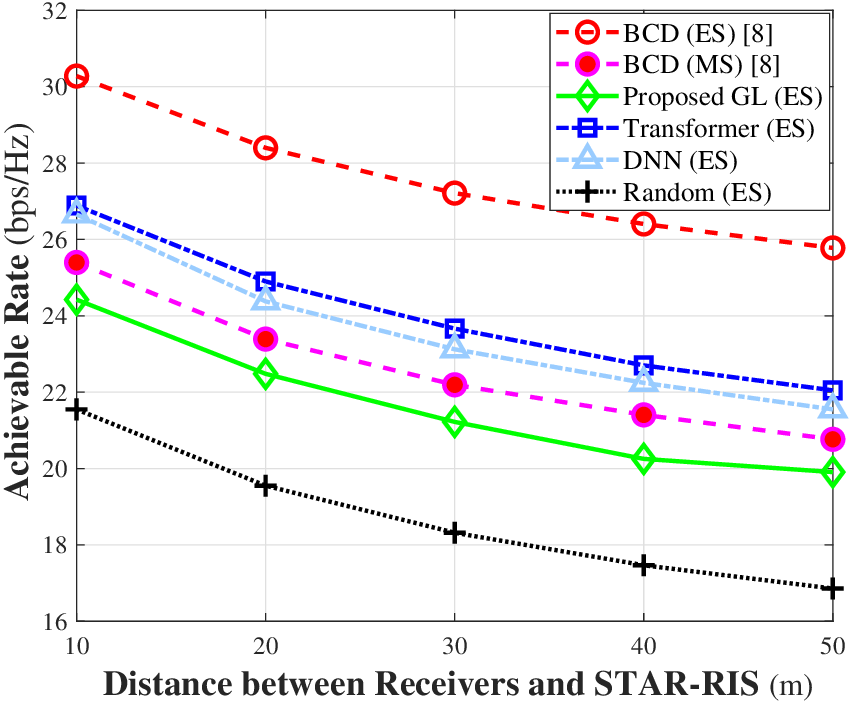}
    \end{subfigure}
    \captionsetup{font=small} 
    \caption{Achievable rate versus number of RIS elements under different precoding schemes ($M=8$, $N_r=4$).}\label{fig:element} 
\vspace{-0.1in}
\end{figure}

Fig.~\ref{fig:element} shows the impact of RIS size on the achievable rate. As the number of RIS elements increases from $16$ to $54$, all methods exhibit performance gains due to enhanced beamforming capability. GL scales consistently with RIS size, achieving an average rate within $18\%$ of the BCD (ES) solution. Compared to the DNN and Transformer, GL maintains a rate gap below $14\%$, demonstrating strong generalization to spatially expanded scenarios.

\begin{figure}[t]
    \centering
    \begin{subfigure}[b]{0.31\textwidth}
        \includegraphics[width=\textwidth]{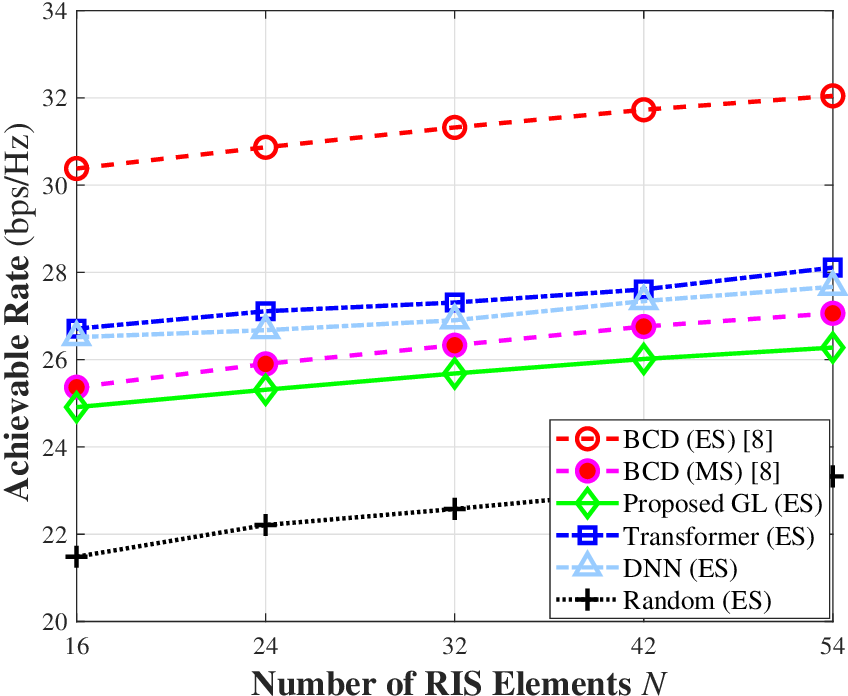}
    \end{subfigure}
    \captionsetup{font=small} 
    \caption{Achievable rate versus user distance to the STAR-RIS under different precoding schemes ($M=8$, $N_r=4$).}\label{fig:distance} 
\vspace{-0.2in}
\end{figure}

Fig.~\ref{fig:distance} illustrates how the achievable rate varies as the user moves further from the STAR-RIS. Although the absolute rate declines for all methods due to increased path loss, GL maintains a stable gap of approximately $21.7\%$ from the BCD (ES) benchmark and tracks the Transformer and DNN within $13\%$ on average. This robustness across a $10$ to $50$ m range highlights the suitability of GL for dynamic or mobile environments.

\section{Conclusion}
In this paper, we have proposed a GL-based precoding framework for STAR-RIS-assisted mmWave MIMO systems. Unlike conventional BCD methods that require perfect CSI and iterative optimization, the proposed method performs inference directly on uplink pilot signals in a CSI-free manner. The learning pipeline combines subspace approximation, supervised feature selection, and XGBoost regression to jointly predict the transmit precoder and STAR-RIS configuration. While supervised feature selection of GL uses BCD under full CSI, the inference process does not require any channel estimation. Simulation results show that the proposed GL approach achieves competitive spectral efficiency compared to BCD and DL-based models, while reducing FLOPs by over four orders of magnitude. These results highlight GL's potential for practical deployment in energy- and hardware-constrained wireless environments. Future work may explore extensions to multi-user scenarios, coupled phase shift constraints, and extending the GL concept to time-varying channels, which are promising directions toward practical deployment.

\end{document}